# TEMPERATURE DEPENDENCE OF PIEZOELECTRIC AND ELECTROSTRICTIVE PROPERTIES IN A PBN:65 MORPHOTROPIC PHASE BOUNDARY THIN FILM

Xi Yang[1], Andrew Beckwith[2]

1) Fermi National Accelerator Laboratory, P.O. Box 500, Batavia, Illinois 60510-0500

2) TcSAM, Department of Physics, University of Houston

Houston, Texas, 77204-5005

## Abstract

We used an optical reflection method to measure the temperature-dependence of the piezoelectricity, electrostriction and time constant of a morphotropic phase boundary (MPB) thin film. We obtained the piezoelectric constant and the electrostrictive constant as a function of temperature by curve fitting the experimental data. We also obtained the time constant as a function of temperature and applied voltage. An abnormal behavior of the strain that was larger at 275 K than at 285 K was observed. This observation was also consistent with the minimal hysteresis behavior of the time constant at 275 K. We provided a possible explanation for this observation.

**INTRODUCTION**

The ferroelectric lead barium niobate solid-solution system, $Pb_{1-x}Ba_xNb_2O_6$ (PBN) is potentially important for optical and electronic applications because of its unique properties as a tungsten bronze-type relaxor ferroelectric with a morphotropic phase boundary (MPB) that separates the tetragonal ferroelectric phase from the orthorhombic ferroelectric phase[1-5]. For PBN compositions near the MPB, the electro-optic (EO)[6, 7], piezoelectric[3], electrostrictive, and pyroelectric[8] properties are enhanced. The large spontaneous polarization and high, switchable polarization near this boundary are especially significant for guided-wave electro-optic devices, surface acoustic wave (SAW) devices, transducers, sensors, actuators and adaptive structures.

The production of high quality thin films of PBN:65 has been driven by the trend toward miniaturization of electronic components, replacement of expensive single-crystal ferroelectrics, and the flexibility of thin films for integration into different types of devices[9]. The characterization of the piezoelectricity and electrostriction of thin-film materials is very important. The converse effect is generally more often to be used to measure piezoelectric and electrostrictive constants than the direct effect, although the small strains produced by an electric field of a reasonable size can lead to experimental difficulties[10]. In terms of the configuration of the electrodes and the propagation mode of the light, the detection methods can be summarized into the follow two categories:

(1)    the transmission-mode measurement for thin films grown on transparent substrates;

(2)    the reflection-mode measurement for thin films grown on both opaque and transparent substrates.

Under these two categories, an either intensity- or phase- sensitive detection scheme can be adopted[11].

The transmission-mode measurement only can be applied to films deposited on transparent substrates while integration of micro-devices with opaque semiconductor substrates is becoming increasingly desirable. The reflection-mode measurement can be applied to films deposited on either transparent or opaque substrates. Furthermore, in the situation of the temperature-dependent measurement, since the element used to vary the temperature of the sample generally involves an opaque thermal stage, the only possible experiment is the reflection.

**EXPERIMENT AND RESULT**

*Sample Preparation*: A (001) oriented PBN:65 thin film with a thickness of 5000 Å was deposited on MgO substrate using a PLD technique. Interdigital electrodes (IDE) consisting of 10-μm conducting lines separated by 10-μm gaps, as shown in Fig. 1(a), were deposited on top of the PBN thin film. Each line consists of a thin layer of chromium (2500 Å), and external contacts were made to these electrodes using soft indium. A Peltier element served to both cool and heat the sample[12]. The temperature range of the Peltier element is from 242.5 K to 350 K.

*Experimental Setup and Explanation*: A schematic of the reflection experiment is shown in Fig. 1(b). A 48-mm focal length lens was used to focus the incident beam to a spot less than 40 μm in diameter on the surface of the substrate, and a CCD line camera was used to monitor the interference pattern reflected from the substrate[13]. In the experiment, the reflection from two surfaces of the MgO substrate formed an interference pattern. Since during the field-induced bending process, a particular interference peak always corresponded to a constant incident angle, the angle of this interference peak rotating during the bending process was equal to the field-induced bending angle. This method was experimentally examined using the direct reflection technique from an aluminum foil attached to the MgO substrate[14]. The equation

$$E \approx V / b \tag{1}$$

was used to estimate the electric field produced by an applied voltage V. And b was the gap width (10 μm). The equation[13]

$$\theta = \Delta x / d \tag{2}$$

was used to calculate the bending angle. Here, $\Delta x$ was the shift of a particular interference peak recorded by the line camera, and $d$ was the distance between the substrate and the line camera.

*Experimental Result and Explanation*: Considering external contacts made by soft indium, we only cooled the sample instead of heating it, and repeated the same measurement at five different temperature points, they were 255 K, 265 K, 275 K, 285 K and 295 K. At a constant temperature 255 K, hysteresis curves were obtained by recording a data point every 10 s and 300 s with a constant voltage applied to the PBN sample. Each curve in Fig. 2(a) was obtained by varying the applied voltage among 5 volts, 10 volts, 15 volts, and 20 volts. Figs. 2(b), 2(c), 2(d), and 2(e) were obtained at a constant temperature of 265 K, 275 K, 285 K and 295 K separately.

The time constants for switching on each applied voltage and switching off each applied voltage were obtained by curve fitting using the equation

$$\theta(t) = \Delta\theta \cdot [1 - e^{-t/\tau}] + \theta_0 \tag{3}$$

The results are shown in Fig 3. Figs. 3(a), (b), (c), and (d) are expressed in a relationship between the time constant and the temperature when a constant voltage of 5 volts, 10 volts, 15 volts, and 20 volts was switched on and off separately. Each curve in Fig. 3(e) is expressed in a relationship between the time constant and the applied voltage at a constant temperature of 255 K, 265 K, 275 K, 285 K and 295 K. The common property among Figs. 3(a), (b), (c), and (d) is that the difference between time constants of switching on and off a constant voltage had a minimal value at the vicinity of the temperature 275 K.

The relationship between the strain and the applied voltage at a constant temperature of 255 K, 265 K, 275 K, 285 K and 295 K is shown in Fig. 4. And all of the measurement was done when the film was under compression from the MgO substrate. We obtained the piezoelectric constant, the electrostrictive constant, and the residual strain at each constant temperature among 255 K, 265 K, 275 K, 285 K and 295 K by polynomial curve fitting to the second order. The result is shown in Fig. 5. The relationship between the piezoelectric constant and the temperature is shown in Fig. 5(b), the relationship between the electrostrictive constant and the temperature is shown in Fig. 5(c), and the relationship between the residual strain and the temperature is shown in Fig. 5(a). From Figs. 5(b) and 5(c), at 265 K, when the piezoelectric constant had a minimal value, the electrostrictive constant had a maximal value; the situation was reversed at 285 K. The piezoelectricity is coming from parallel and anti-parallel domains (relative to the direction of the external electric field), and the electrostriction is coming from the 90° domains. The possible explanation for the above observation is that 90° domains were dominant over parallel and anti-parallel domains at the temperature 265 K, and parallel and anti-parallel domains were dominant over 90° domains at the temperature 285 K. When the temperature had a value somewhere between 265 K and 275 K, parallel anti-parallel domains coexisted with 90° domains. This also might be able to explain that at an applied voltage of 20 volts the strain was larger at 275 K than at 285 K. This is inconsistent with the general situation that the strain increases with increasing temperature. Also, from Fig. 3, the hysteresis behavior of the time constant was minimized at 275 K when an applied voltage was switched on and off.

## CONCLUSIONS

We used a reflection method[13] developed recently in our lab to measure the temperature dependence of piezoelectric and electrostrictive properties of a morphotropic phase boundary PBN:65 thin film sample. From these measurements, we obtained piezoelectric constants and electrostrictive constants at different temperatures. Also by curve fitting, we obtained the time constant that was varied with temperature and the applied voltage. We observed an abnormal phenomenon that was the strain developed by an applied 20 volts was larger at 275 K than it was at 285 K. At 275 K, hysteresis behavior between time constants of switching on and off an applied voltage was also minimized at 275 K. We explain this abnormal behavior by the coexistance of the parallel anti-parallel and 90° domains. We also provide evidence from the curve fitting result of the piezoelectricity and electrostriction.

*Acknowledgments*

The authors thank Dr. Lowell T. Wood for providing optical facilities, Dr. Markus Aspelmeyer for providing the Peltier element, and Dr. Mikhail Strikovski for preparing the thin film. The authors acknowledge support from the Robert A. Welch Foundation (E-1221).

Figure Captions

FIGURE 1   (a) Interdigital planar electrodes (IDE); (b) Schematic diagram of the reflection experiment on the PBN sample.

FIGURE 2   Strain-electric field hysteresis curves.  They were obtained by recording a data point every 10 s for 300 s with a constant voltage applied to the PBN sample.  (a) at a constant temperature 255 K; (b) at a constant temperature 265 K; (c) at a constant temperature 275 K; (d) at a constant temperature 285 K; (d) at a constant temperature 295 K.

FIGURE 3   Time constant-temperature curves.  At each applied constant voltage, the solid line shows the time constant-temperature relationship when the constant voltage was switched on and the dashed line shows the time constant-temperature relationship when the constant voltage was switched off.  (a) at a constant voltage of 5 volts; (b) at a constant voltage of 10 volts; (c) at a constant voltage of 15 voltages; (d) at a constant voltage of 20 volts; (e) time constant-applied voltage relationship, each curve at a constant temperature of  255 K, 265 K, 275 K, 285 K, and 295 K.

FIGURE 4   Strain-applied voltage relationship.  Each curve at a constant temperature of 255 K, 265 K, 275 K, 285 K, and 295 K.

FIGURE 5   (a) Residual strain-temperature relationship; (b) piezoelectric constant-temperature relationship; (c) electrostrictive constant-temperature relationship.

(a)

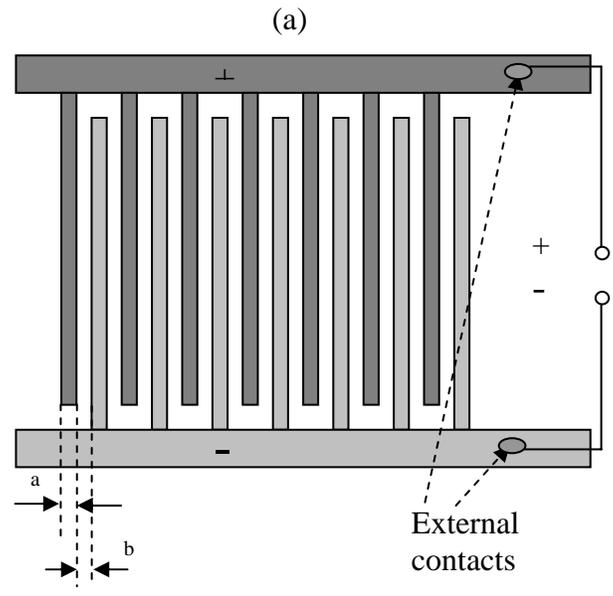

(b)

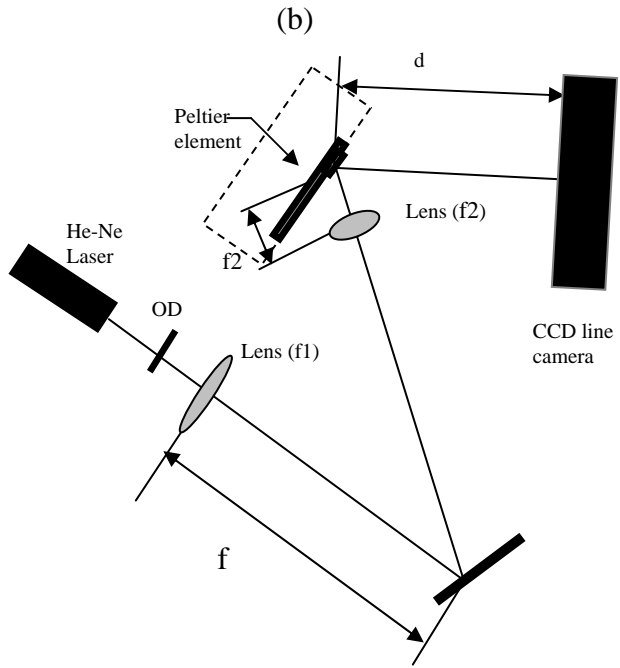

Fig. 1

Yang et al.

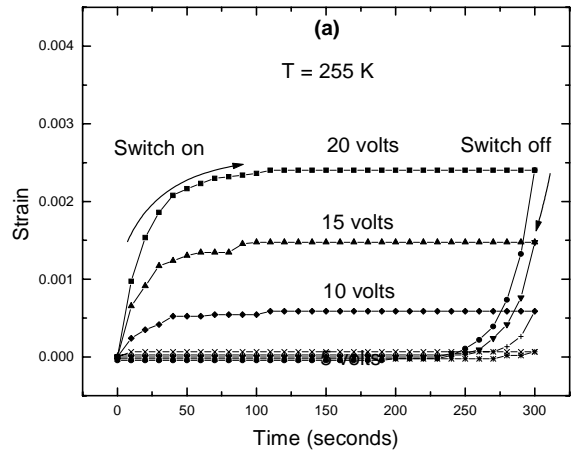

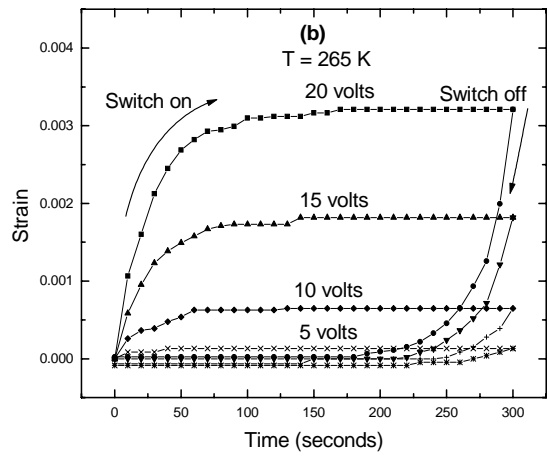

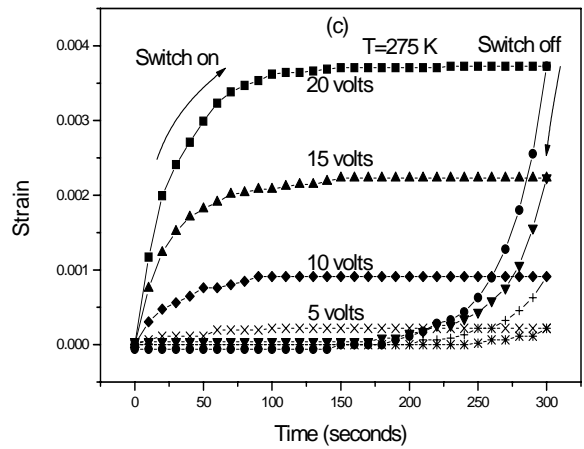

Fig. 2
Yang et al

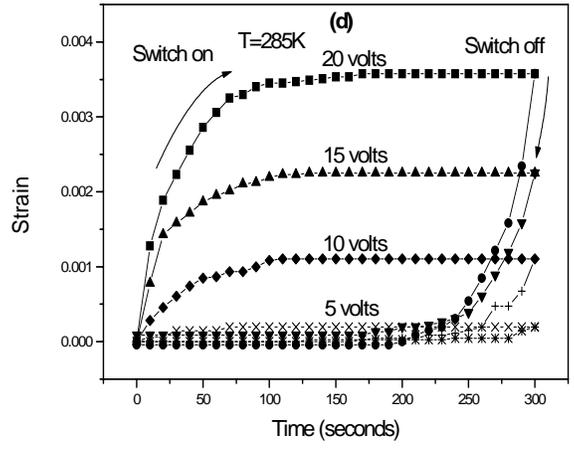

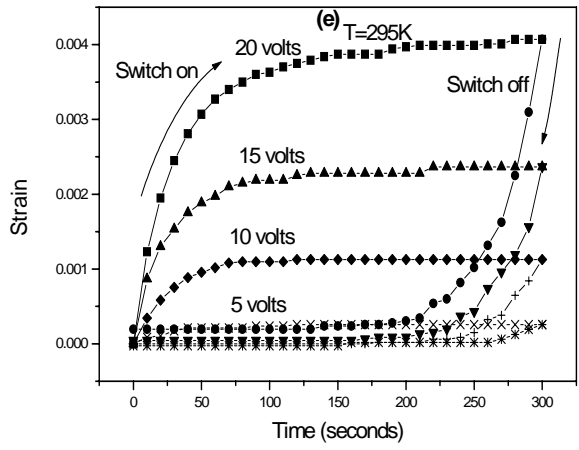

Fig. 2
Yang et al

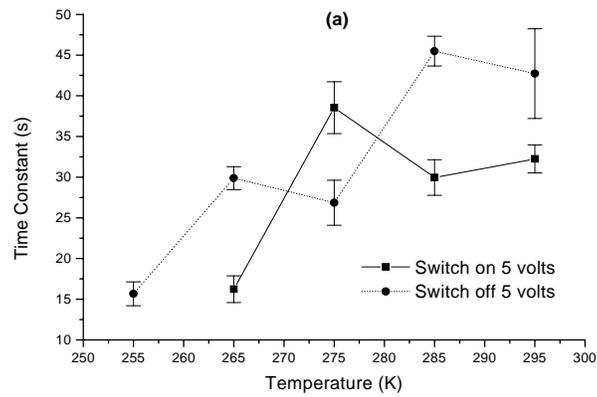

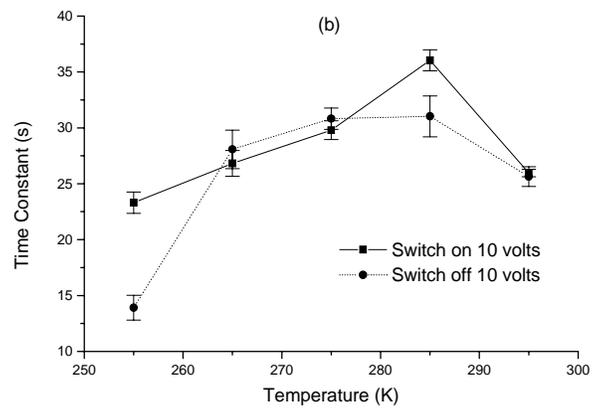

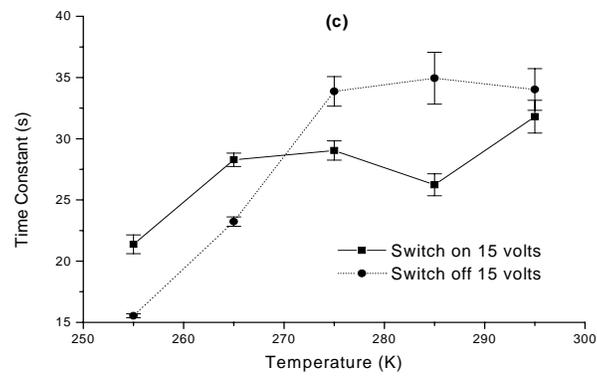

Fig. 3
Yang et al

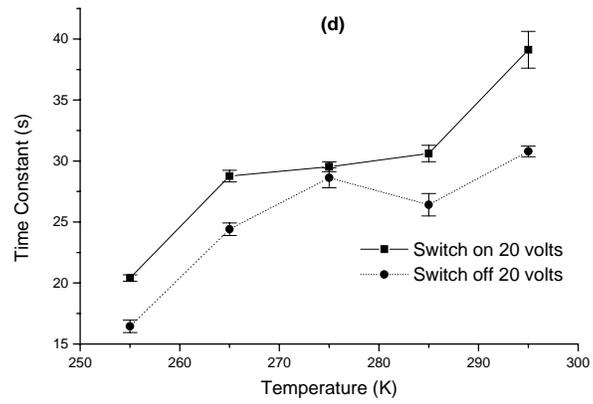
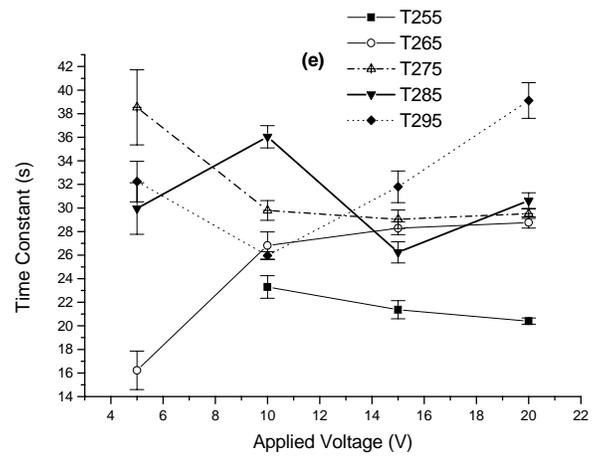

Fig. 3
Yang et al

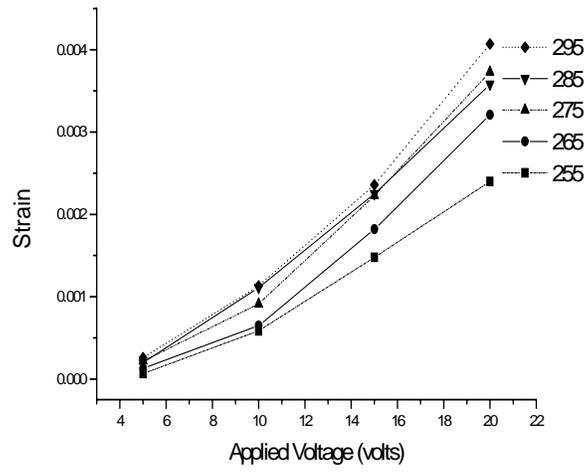

Fig. 4
Yang et al

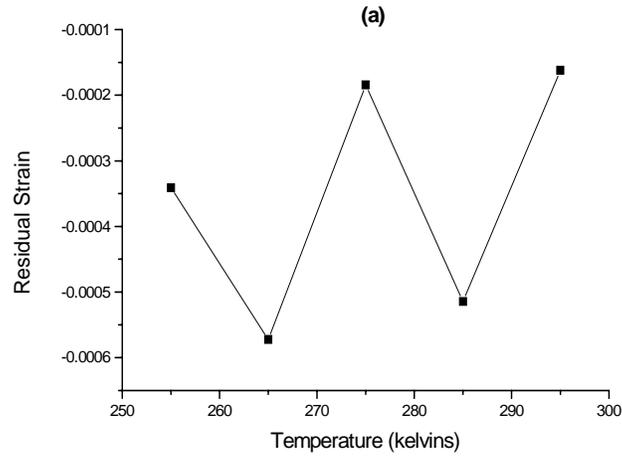

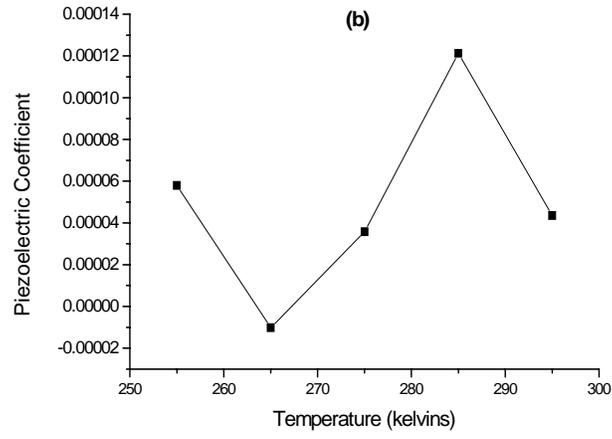

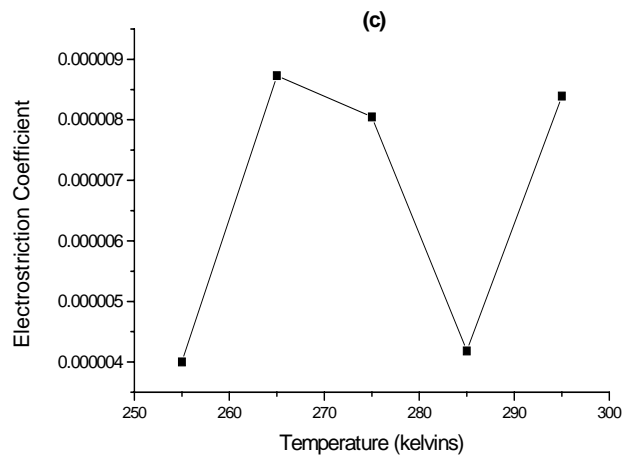

Fig. 5
Yang et al

16